# TOWARDS FUZZY-HARD CLUSTERING MAPPING PROCESSES

**MINYAR SASSI**
*National Engineering School of Tunis*
*BP. 37, Le Belvédère, 1002 Tunis, Tunisia*

Although the validation step can appear crucial in the case of clustering adopting fuzzy approaches, the problem of the partition validity obtained by those adopting the hard ones was not tackled. To cure this problem, we propose in this paper fuzzy-hard mapping processes of clustering while benefitting from those adopting the fuzzy case. These mapping processes concern: (1) local and global clustering evaluation measures: the first for the detection of the "worst" clusters to merging or splitting them. The second relates to the evaluation of the obtained partition for each iteration, (2) merging and splitting processes taking into account the proposed measures, and (3) automatic clustering algorithms implementing these new concepts.

## 1. Introduction

Classification problem is a process for grouping a set of data into groups so that data within a group have high similarity, but are very dissimilar to data in other groups.

This problem is declined in two alternatives: supervised [1,2] and unsupervised approaches [3].

In the first, we know the possible groups and we have a data already classified, being used overall as training. The problem consists in associating the data in the most adapted group while being useful of those already labeled.

In unsupervised classification, also called clustering, possible groups (or clusters) are not known in advance, and the data available are not classified. The goal is then to classify in the same cluster the data considered as similar.

Since clustering is an unsupervised method, there is a need of some kind of clustering result quality validation.

This quality is judged in general on the basis of two contradictory criteria [4]. The first supposes that generated clusters must be most various possible to each other, and the second requires that each cluster have to be most homogeneous possible.

The data are grouped into clusters based on a number of different approaches. Partition-based clustering and hierarchical clustering are two of the main techniques.

Hierarchical clustering techniques [5] generate a nested series of partitions based on a criterion, which measures the similarity between clusters or the separability of a cluster, for merging or splitting clusters. We can mention, BIRCH (Balanced Iterative Reducing and Clustering using Hierarchies) algorithm [6], CURE (Clustering Using REpresentatives) algorithm [5].

Partition-based clustering often starts from an initial partition and optimizes (usually locally) a clustering criterion. A widespread accepted clustering scheme subdivides these techniques in two main groups: hard and fuzzy [3]. The difference between these is mainly the degree of membership of each data to the clusters. During the construction of the clusters, in the case of the hard group, each data belongs to only one cluster with a unitary degree of membership, whereas, for the fuzzy group, each data can belong to several clusters with different degrees of membership. We mention some algorithm like k-means [7] and the Fuzzy C-Means (FCM) [8]. In this work we limit ourselves to partition-based clustering methods.

Since most basic clustering algorithms assume that the number of clusters is a user-defined parameter, which is difficult to known in advance in real applications. Thus it is difficult to guarantee that the clustering result can reflect the natural cluster structure of the datasets. Several work tackled this problem [9,10,11].

When we are confronted with a problem of determination of the number of clusters, we brought to make assumptions on this last.

To prevent user to choose this number, a solution consists in making iterations until obtaining an optimal number of clusters. Each iteration tries to minimize (or maximize) an objective function called validity index [8,12,13,14] which measures the clustering quality to choose the optimal partition among all those obtained with the various plausible values of the required cluster's number.

In [11,15], clustering approaches adopting the fuzzy concept were presented and proven. They are based on merging and splitting processes.

Although these processes can appear crucial in the fuzzy case, the problem of the validity of the partition obtained by automatic hard clustering methods wasn't tackled.

To cure this problem, we propose in this article rules for mapping clustering for hard partition-based clustering while benefitting from those adopting the fuzzy partition-based approaches.

These mapping rules concern:

- The definition of local and global measures for clustering evaluation: the first for the detection of worst clusters to merging or splitting them with others. The second relate to the evaluation of the obtained partition in each iteration.
- The binarization of merging and splitting processes.
- The modeling and implementation of automatic hard clustering algorithm taking into account the new concepts.

Towards Fuzzy-Hard Clustering Mapping Processes

The rest of the article is organized as follows. In section 2, we discuss backgrounds related to fuzzy and hard clustering techniques. In section 3, we present our motivation. In section 4, we present the mapping fuzzy-hard processes. Section 5 gives the experimentation and finally, section 7 concludes the paper and gives some futures works.

## 2. Backgrounds

Clustering methods group homogeneous data in clusters by maximizing the similarity of the objects in the same cluster while minimizing it for the objects in different clusters.

To make it easier for the readers understand the ideas clustering techniques, we tried to unify notations used in this last.

To achieve that, the following definitions are assumed: $X \in R^{N \times M}$ denotes a set of data items representing a set of $N$ objects $x_i$ in $R^M$, $c_j$ denoted the $j^{th}$ cluster and $c$ denoted the optimal number of clusters found.

In this section, we present basic concepts related to fuzzy and hard clustering.

### 2.1. *Fuzzy Clustering*

Fuzzy clustering methods allow objects to belong to several clusters simultaneously, with different degrees of membership [16]. The data set $x$ is thus partitioned into $c$ fuzzy subsets [17]. The result is the partition matrix $U = \lfloor \mu_{ji} \rfloor$ for $1 \le i \le N, \ 1 \le j \le c$.

Several researches were carried out for the automatic determination of the number of clusters and the quality evaluation [13,15,11] of the obtained partitions.

Bezdek [18] introduced a family of algorithms known under the name of Fuzzy C-Means (FCM).

Te objective function $(J_m)$ minimized by FCM is defined as follows:

$$J_m(U,V) = \sum_{j=1}^{c} \sum_{i=1}^{N} \mu_{ji}^m \left\| x_i - c_j \right\|^2$$

$U$ and $V$ can be calculated as :

$$\mu_{ji} = \frac{\left\| x_i - c_j \right\|^{(\frac{1}{1-m})^2}}{\sum_{j=1}^{c} \left\| x_i - c_j \right\|^{\frac{1}{1-m}}}, \ c_j = \frac{\sum_{i=1}^{N} (\mu_{ji})^m x_i}{\sum_{i=1}^{N} (\mu_{ji})^m}$$

where,

$\mu_{ji}$ is the membership value of the $i^{th}$ example, $x_i$, in the $j^{th}$ cluster,

$C = \{c_1, c_2, ..., c_c\}$ a cluster scheme $C' = \{ c_{pk} | C$ and $C_{pk}$ is not a singleton, $k = 1,2,...,m$ where $m = |C'|$.

FCM clustering algorithm is as follows :

---

Algorithm 1 FCM algorithm

---

Inputs: The dataset $X = \{x_i : i = 1,...,N\} \subset R^M$, the number of clusters $c$, the fuzzifier parameter $m$ and Euclidian distance function $\|\ \|$.

Outputs: The cluster centers $c_j^1 (j = 1,2,...,c)$, the membership matrix $U$ and the elements of each cluster $i$, i.e., all the $x_i$ such that $u_{ji} > u_{ki}$ for all $k \neq j$.

Step 1: Input the number of clusters $c$, the fuzziffer $m$ and the distance function.

Step 2: Initialize the cluster centers $c_j^0 (j = 1,2,...,c)$

Step 3: Calculate $u_{ji} (i = 1,2,...,N; j = 1,2,...,c)$.

Step 4: Calculate $c_j^1 (j = 1,2,...,c)$.

Step 5: If $\max_{1 \leq j \leq c} \left( \|c_j^0 - c_k^1\| / \|c_k^1\| \right) \leq \varepsilon$ then go to step 6; else let $c_j^0 = c_j^1 (j = 1,2,...,c)$ and go to step 3.

---

A simpler way to prevent bad clustering due to inadequate cluster's centers is to modify the basic FCM algorithm. We start with a large (resp. small) number of uniformly distributed seeds in the bounded $M$-dimensional feature space, but we decrease (resp. increase) them considerably by merging (resp. splitting) worst clusters until the quality measure stops increasing (resp. decreasing).

In consequent, compactness and separation are two reasonable measures making it possible to evaluate the quality of obtained clusters.

Automatic fuzzy clustering algorithm is based on global measure representing the separation-compactness for clustering quality evaluation.

Given a cluster scheme $C = \{c_1, c_2, ..., c_c\}$ for a dataset $X = \{x_1, x_2, ..., x_N\}$, let $c' = \{ c_{pk} | C$ and $C_{pk}$ is not singleton, $k = 1,2,...,m$ where $m = |C'| \}$, the global separation-compactness, $SC$, of a cluster scheme $C$ is given by:

Towards Fuzzy-Hard Clustering Mapping Processes

$$SC = \left( \frac{k}{c} \times \frac{\sum_{j=1}^{c} \min_{1 \leq k \leq c} \left\{ \left\| v_j - c_k \right\|^2 \right\}}{\sum_{j=1}^{k} \left( \sum_{x_i \in C_{pj}, x_i \neq v_j} \mu_j(x_i)^2 \left\| x_i - c_j \right\|^2 / \sum_{x_i \in C_{pj}, x_i \neq v_j} \mu_j(x_i)^2 \right)} \right)$$

where $u_j(x_i)$ is the membership value of $x_i$ belonging to $j^{th}$ cluster. $v_j$ is the center of $j^{th}$ cluster $c_j$. $c$ is the number of clusters and $2 \leq c \leq \sqrt{N}$.

While basing on these measures, merging (resp. splitting) clustering algorithm is as follows:

| Algorithm 2 Iteratif clustering process |
| --- |

Inputs: A dataset $X = \{x_i : i = 1, ..., N\} \subset R^M$ and the Euclidian distance function $\| \|$.

Outputs: Output $C = \{c_1, c_2, ..., c_{opt}\}$ as an optimal cluster centers.

Step 1: Initialize the parameters related to the FCM, $c = c_{max}$, $c_{min} = 2$, $c_{opt} = \sqrt{N}$,

Step 2: Apply the basic FCM algorithm to update the membership matrix $(U)$ and the cluster schema.

Step 3: Test for convergence; if no, go to step 2.

Step 4: Calculate $SC$ measure.

Step 5: Repeat

Perform merging (resp. splitting) process to get candidate cluster centers, decrease $c \leftarrow c - 1$ (resp. increase $c \leftarrow c + 1$); perform the basis FCM algorithm based on parameter $c$ to find the optimal cluster centers. Calculate $SC$ measure for new clusters, denote it as $SC'$; if $SC' > SC$ then $SC = SC'$, $c_{opt} = c$.

Until $c \leq 2$.

### 2.1.1. *Fuzzy Merging Process*

The fuzzy merging process generally used by earlier studies involves some similarity or compatibility measure to choose the most similar or compatible pair of clusters to merge into one. In our merge process, we choose the "worst" cluster and delete it. Each element included in this cluster will then be placed into its own nearest cluster. Then, centers of all clusters will be adjusted. That means, our merge process may affect multiple clusters, which we consider to be more practical. How to choose the "worst" cluster? We still use the measures of separation and compactness to evaluate individual clusters (except singleton).

Given a cluster scheme $C = \{c_1, c_2, ..., c_c\}$ for a dataset $X = \{x_1, x_2, ..., x_N\}$, for each $c_j \in C$, if $c_j$ is not singleton, the local separation-compactness of $c_j$, denoted as $sc_j$, is given by:

$$sc_j = \min_{1 \le j \le c, k \ne j} \|c_j - c_k\|^2 \times \frac{\sum\limits_{x_i \in C_j, x_i \ne v_j} u_j(x_i)^2}{\sum\limits_{x_i \in C_j, x_i \ne v_j} u_j(x_i)^2 \|x_i - c_j\|^2}$$

where $u_j(x_i)$ is the membership value of $x_i$ belonging to $j^{th}$ cluster. $c_j$ is the center of $j^{th}$ cluster $c_j$. $c$ is the number of clusters and $2 \le c \le \sqrt{N}$.

A small value of $sc_j$ indicates the "worst" cluster to be merged.

### 2.1.2. *Fuzzy Splitting Process*

In this process, we operate by splitting the "worst" cluster at each stage in testing the number of clusters $c$ from $c_{min}$ to $c_{max}$.

The global separation-compactness measure is used.

The general strategy adopted for the new algorithm is as follows: at each step of the algorithm, we identify the "worst" cluster and split it into two clusters while keeping the other $c-1$ clusters.

The general idea in the splitting approach is to identify the "worst" cluster and split it, thus increasing the value of $c$ by one. Our major contribution lies in the definition of the criterion for identifying the "worst" cluster.

For identifying the "worst" cluster, a "score" function $s(j)$ associated with each cluster $j$, as follows:

$$s(j) = \frac{\sum_{i=1}^{N} \mu_{ij}}{number\_of\_data\_vectors\_in\_cluster\_j}$$

In general, when $s(j)$ small, cluster is $j$ tends to contain a large number of data vectors with low membership values.

The lower the membership value, the farther the object is from its cluster center. Therefore, a small $s(j)$ means that cluster $j$ is large in volume and sparse in distribution. This is the reason that the cluster corresponding to the minimum of $s(j)$ as the candidate to split when the value of $c$ is increased.

On the other hand, a larger $s(j)$ tends to mean that cluster $j$ has a smaller number of elements and exerts a strong "attraction" on them.

## 2.2. *Hard Clustering*

In hard clustering, the data can be gathered in a table with $N$ lines and $M$ columns. If the data belong to a set of clusters, it is possible to associate to this data table a membership table whose values 1 or 0 respectively imply the membership of cluster $C_j$, with $j = 1, 2, ..., c$.

One of most known hard clustering algorithms is k-means. Its goal is to minimize the distance from each data compared to the cluster center to which it belongs. This algorithm corresponds to the search of centers $c_j$ minimizing the following criterion: [19]

$$E = \frac{1}{2} \sum_{j=1}^{c} \sum_{i=1}^{N} \delta_{ji} \left\| x_i - c_j \right\|^2$$

where

$$\delta_{ji} = \begin{cases} 1 \, if \, x_i \in c_j \\ 0 \, else \end{cases}$$

k-means algorithm consists in choosing initial centers and improving the partition obtained in an iterative way in three steps.

---

Algorithm 3 k-means algorithm

---

Inputs: A dataset $X = \{x_i : i = 1, ..., N\} \subset R^M$ and the Euclidian distance function $\| \, \|$.

Output: Cluster scheme $C = \{c_j : j = 1, ..., c\} \subset R^M$.

Step 1: Initialize the $c$ centers with randomly values.

Step 2: Affect each data in the nearest cluster $C_j$ : $x_i \in c_j$, if $\left\| x_i - c_j \right\| \prec \left\| x_i - c_l \right\|$ for $i = 1 ... N$, $l \neq j$, $j = 1 ... c$.

Step 3: Recalculate the position of each new center : $c_j^* = \frac{1}{N_j} \sum_{x_i \in c_j} x_i$, where $N_j$ the cardinality of center $c_j$ and

$j = 1, ..., c$.

---

$N$ and $k$ are the size of data and the number of clusters respectively. It is necessary to repeat steps 2 and 3 until convergence, i.e. until the centers do not change.

The result of implementation of this algorithm does not depend on the data order input. It has a linear complexity [19] and adapts to the large data sets and requires fixing the number of clusters which influences the output. The result is sensitive to the starting situation, as well on the number of cluster as the initial center's positions.

When we are confronted with a problem of determination of the number of clusters, we brought to make assumptions on this last.

To prevent user to choose this number, a solution consists in making iterations until obtaining an optimal number of clusters. Each iteration tries to minimize (or maximize) an objective function called validity index [8,12,13,14] which measures the clustering quality to choose the optimal partition among all those obtained with the various plausible values of the required cluster's number.

In this sub-section, we present, firstly, some validity indices; then, we give new definitions of separation and compactness quality measures applicable in the context of hard partition-based clustering.

The Mean Squared Error (MSE) [20] of an estimator is one of many ways to quantify the amount by which an estimator differs from the true value of the quantity being estimated. MSE measures the average of the square of the error. The error is the amount by which the estimator differs from the quantity to be estimated. The difference occurs because of randomness or because the estimator doesn't account for information that could produce a more accurate estimate. In clustering, this measure corresponds to compactness.

$$MSE = \frac{1}{N} \sum_{j=1}^{c} \sum_{i=1}^{N} \delta_{ji} \left\| x_i - c_j \right\|^2$$

In the case of a hard partition-based clustering, the Dunn index [21] takes into account the compactness and the separability of the clusters: the value of this index is all the more low since the clusters are compact and quite separate. Let us note that the complexity of the index of Dunn becomes prohibitory as soon as large data sets; it is consequently seldom used.

$$I_{Dum} = \frac{\min\{D_{\min}(c_j, c_k)\}}{\max\{D_{\max}(c_j)\}}$$

where $D_{\min}(c_j, c_k)$ is the minimal distance which separates a data from cluster $c_j$ to data from cluster $c_k$ and $D_{\max}(c_j)$ is the maximal distance which separates two data from cluster $c_j$ :

$$D_{\min}(c_j, c_k) = \min\{\left\| x_i - y_i \right\| : x_i \in c_j \, et \, y_i \in c_k\}$$

$$D_{\max}(c_j) = \max\{\left\| x_i - y_i \right\| : (x_i, y_i) \in c_j\}$$

Towards Fuzzy-Hard Clustering Mapping Processes

The Davies-Bouldin index [22] of the clustering is combined with the centroid diameters comparison between clusters. In the computation of the Davies-Bouldin index, the centroid linkage is used as the inter-cluster distance. The centroid inter-cluster and intra-cluster measures are selected for compatibility with the k-means clustering algorithm used in the sensors (which essentially computes centroids of clusters at each iteration).

This index takes into account the compactness and the separability of the clusters: its value is all the more low since the clusters are compact and quite separate. It is defined by the following expression:

$$I_{DB} = \frac{1}{c} \sum_{j=1}^{c} \max_{j \neq l} \frac{\{D_c(c_j) + D_c(c_k)\}}{D_{ce}(c_j, c_k)}$$

where $D_c(c_j)$ is the average distance between a data from cluster $c_j$ and his center, $D_{ce}(c_j, c_k)$ is the distance which separates the cluster's centers $c_j$ and $c_k$. It is defined by the following expression:

$$D_c(c_j) = \frac{1}{N} \sum_{i=1}^{N} \|x_i - c_j\|$$

$$D_{ce}(c_j, c_k) = \|c_j - c_k\|$$

## 3. Motivations

When we are confronted with a problem of determination of the number of clusters, we brought to make assumptions on this last.

To prevent user to choose this number, a solution consists in making iterations until obtaining an optimal number of clusters. Each iteration tries to minimize (or maximize) an objective function called validity index which measures the clustering quality to choose the optimal partition among all those obtained with the various plausible values of the required cluster's number.

The previously presented indices are applicable in the case of hard partition-based clustering and prove their limits in determination of optimal number of clusters.

Moreover, all these measures are global, i.e. they allow the evaluation of all the partition.

However, in the case of automatic clustering where the number of clusters is unknown for the user, we will need to iterate a clustering algorithm until obtaining the optimal number of clusters. Therefore we will have recourse to local evaluation for the detection of worst clusters to be merged or split.

While basing on these principles, we propose to evaluate, globally and locally, the obtained to reach the optimum. The global evaluation makes it possible to judge on the quality of generated partition whereas the local evaluation makes it possible to detect the worst clusters.

## 4. Mapping Processes

When we are confronted with a problem of determination of the number of clusters, we brought to make assumptions on this last.

As mentioned in section 2, clustering algorithms allow to partition data in clusters by maximizing the similarity of the data in the same cluster while minimizing it for data in the various clusters. Consequently, compactness and separation are the two reasonable criteria allowing the cluster's quality evaluation. The proposed quality evaluation measures are based on these measures.

Global evaluation makes it possible to judge on the quality of generated partition. We propose here new formulations which define hard global compactness and hard global separation within the partition.

**Hard Global Compactness.** Given a cluster scheme $C = \{C_1, C_2, ..., C_c\}$ for a data set $X = \{x_1, x_2, ..., x_N\}$, let $C^{'} = \{ C_j | c$ and $C_j$ is not a singleton, $j = 1, 2, ..., k$ with $k = |C^{'}| \}$, the hard global compactness, $Cmp$, of the cluster scheme $C$ is given by:

$$Cmp = \frac{k}{\sum_{j=1}^{k} Var_j}$$

where $Var_j$ is the variance of $j^{th}$ cluster. It's given by the following expression:

$$Var_j = \frac{\sum_{x_i \in C_j} \|x_i - c_j\|^2}{card(C_j)^2}$$

where $c_j$ is the center of cluster $C_j$.

**Hard Global Separation.** The hard global separation, Sep, of a cluster scheme $C = \{C_1, C_2, ..., C_c\}$ for a dataset $X = \{x_1, x_2, ..., x_N\}$ is given by:

$$Sep = \left( \frac{\sum_{j=1}^{c} \min\left( \|c_j - c_k\|^2 \right)}{c} \right)^2$$

Towards Fuzzy-Hard Clustering Mapping Processes

**Hard Global Separation-Compactness.** Given a scheme $C = \{C_1, C_2, ..., C_c\}$ for a dataset $X = \{x_1, x_2, ..., x_N\}$, let $C^{'} = \{C_j | C$ and $C_j$ is not a singleton, $j = 1,2,...,k$ where $k = |C^{'}|\}$, the hard global separation-compactness, $SepCmp$, of a cluster scheme $C$, is given by:

$$SepCmp = \frac{k}{c} Sep \times Cmp$$

In consequent, the choice of the best partition is obtained by maximizing the measure $SepCmp$.

For the determination of the optimal number of clusters, we propose two approaches. The first is based on the merging principle, which we called EMk-means (Enhanced Merging k-means). We start with a maximum number of clusters $c_{max}$ and which it decreases during various iterations by identifying the "worst" cluster to merge it.

The second is based on the splitting principle which we called ESk-means (Enhanced Splitting k-means). We start with a minimum number of clusters $c_{min}$ and which it increase during iterations. This increasing is done by dividing the clusters having a maximum value of variance.

The principle of the two adopted approaches is summarized in the following algorithm:

| Algorithm 3 Iterative Clustering Approach |
| --- |

Step 1: Initialize $c_{max}$ and $c_{min}$, Initialize the cluster's centers.

Step 2: Apply k-means algorithm.

Step 3: Calculate $SepCmp$.

Step 4: For $k$ from $c_{max}$ to $c_{min}$ (respectively $c_{min}$ to $c_{max}$) do Merging (respectively splitting) the clusters.

Step 5: Calculate $SepCmp$.

Step 6: Determine the optimal value of the number of cluster $c_{opt}$ whose value $SepCmp$ is maximal.

## 4.1. *Merging Process Mapping*

The local evaluation is used to identify the worst clusters to be merged with the others. Each data belonging to this cluster is affected to the nearest one. Then, the centers of all clusters are adjusted. To identify worst cluster in each iteration, separation and compactness measures are used for local evaluation.

While basing on hard local separation-compactness measures, we present in this sub section mapping rules from fuzzy to hard.

**Hard local Compactness.** Given a cluster scheme $C = \{C_1, C_2, ..., C_c\}$ for a dataset $X = \{x_1, x_2, ..., x_N\}$, for each $c_{j \in C}$, if $c_j$ is not a singleton, the hard local compactness of $C_j$, denoted as $cmp_j$, is given by :

$$cmp_j = \frac{Var(C_j)^2}{\sum_{x_i \in C_j} \|x_i - c_j\|^2}$$

where $Var_j$ is the variance of jth cluster.

**Hard Local Separation.** Given a cluster scheme $C = \{C_1, C_2, ..., C_c\}$ for a dataset $X = \{x_1, x_2, ..., x_N\}$, the hard local separation, denoted as $sep_j$, is given by:

$$sep_j = \min_{\substack{1 \leq j \leq c \\ j \neq l}} \|c_j - c_l\|^2$$

where $c_j$ and $c_l$ is the centers of clusters $C_j$ and $C_l$ respectively.

**Hard Local Separation-Compactness.** Given a cluster scheme $C = \{C_1, C_2, ..., C_c\}$ for a dataset $X = \{x_1, x_2, ..., x_N\}$, for each $C_{j \in c}$, if $C_j$ is not a singleton, the hard local separation-compactness of $C_j$, denoted as $sepcmp_j$, is given by:

$$sepcmp_j = sep_j \times cmp_j$$

Thus, the "worst" cluster is the one with the least $sepcmp_j$ value.

By combining these measures with the k-means algorithm, the number of clusters is thus obtained automatically.

The proposed algorithm, called EMk-means, is based on merge strategy. We start with a maximum number of clusters $c_{max}$ and which it decreases during various iterations by identifying the "worst" cluster to merge it.

For $c_{max}$, we adopted the following Bezdek suggestion [18]: $c_{max} = \sqrt{N}$ (N is the size of the dataset).

In each iteration, the algorithm found to maximize the hard global separation-compactness measure $SepCmp$, obtained for the various plausible values of the required number of clusters.

The worst cluster is identified and merged with the remaining clusters.

The EMk-means algorithm is described below.

---

Algorithm 4 EMk-means algorithm

---

Inputs: A dataset $X = \{x_1, x_2, ..., x_N\}$, the Euclidien distance function $\| \, \|$.

Output: Optimal cluster scheme $C = \{C_1, C_2, ..., C_c\}$.

Step 1: Initialize the parameters related to k-means algorithm $c_{max} = c$, $c_{min} = 2$.

Step 2: Apply the k-means algorithm.

---

Step 3: If convergence, then go to step 3, else go to step 2.

Step 4: Calculate $SepCmp$ measure.

Step 5: Repeat

        Apply merge-based procedure for obtain the candidate cluster scheme.

        Decrease the cluster number $c \leftarrow c - 1$.

        Calculate the $SepCmp$ measure for the new clusters.

    Until $c \leq 2$

Step 6: Store $SepCmp$ measure which value is maximal.

Hard merge-based procedure is presented as follows:

Procedure  Hard merge-based procedure

Input: Cluster scheme $C^* = \{C_1, C_2, ..., C_c\}$

Output: Candidate cluster scheme $C^* = \{C_1, C_2, ..., C_{c-1}\}$

Step 1: Calculate $sepcmp_j$ measure for each cluster belonging to $c^*$.

Step 2: Delete the worst cluster which have the minimal value of $sepcmp_j$.

Step 3: Assign the data of this cluster to the various remaining clusters

Step 4: Calculate the values of the new clusters centers according to the median formula.

Step 5: Apply the k-means algorithm to the new clusters.

## 4.2. *Splitting Process Mapping*

The proposed algorithm is based on a construction strategy, i.e., initialized with a minimum number of clusters $c_{min}$, which is incremented during various algorithm iterations according to a splitting  process until obtaining the maximum cluster number $c_{max} = \sqrt{N}$ , (N is the size of the dataset).

In each iteration, $SepCmp$ measure is calculated for the determination of the optimal number of clusters $c_{opt}$.

In this strategy, we select the worst cluster to divide it into two new clusters.  Each data belonging to this cluster is then affected in a new cluster of which we must again calculate the center.

The splitting-based process is carried out by calculating the variance of each cluster which is given by the following equation:

$$Var\left(C_j\right) = \frac{1}{N_{C_j}} \sum_{x_i \in C_j} D^2\left(x_i - c_j\right)$$

where,

$N_{C_j}$ : the number of data belonging to the cluster $C_j$ .

$D(x_i - c_j)$

: the Euclidian distance between $x_i$ data belonging to the cluster $C_j$ and his center $c_j$

.

The objective of the ESk-means algorithm aims at minimizing the average intra-cluster distance, this is why we choose the cluster corresponding to the maximum value of $v(c_j)$ as a candidate for the splitting-based process.

When the value of $Var(c_j)$ is high, the data belonging to cluster $C_j$ are dispersed, i.e. the data tend to move away from center $c_j$ of cluster $C_j$.

The splitting-based procedure consists on identifying the worst cluster having the maximum value of variance, the set of its data is noted $E$, then calculate for each one as of the its data the total distance compared to the various centers of the not selected classes. The first two data having a maximum total distance of the centers are selected like initial centers being used for starting of the k-means algorithm in order to partitioned $E$ in two new sets $E_1$ and $E_2$.

The binary procedure of division is detailed like following.

---
Procedure  Hard splitting-based procedure

---
Input: Cluster scheme $C^* = \{C_1, C_2, ..., C_c\}$

Output: Candidate cluster scheme $C^* = \{C_1, C_2, ..., C_c, C_{c+1}\}$.

Step 1: Identify the cluster to be divided while calculating its variance. Note $E$ the set of data belonging to this cluster and its center $c_{j0}$.

Step 2: Seek in $E$ two data vectors which their total distances separating them from all the cluster $E$ remainders is maximum. Note these two vectors $c_{j1}$ and $c_{j2}$.

Step 3: Apply the k-means algorithm to new centers $c_{j1}$ and $c_{j2}$. in order to obtain two new partitions $E_1$ and $E_2$.

---

The aim of algorithm is to select the worst cluster, to remove it and to replace it by two other new clusters and this by applying splitting-based process. At the end of each iteration, $SepCmp$ measure is calculated and the clusters number is incremented until obtaining $c_{max}$. This number is allowed to the cluster which value of $SepCmp$ is maximum.

The proposed algorithm for the determination of the values of the centers and $c_{opt}$ clusters is described below:

---

Algorithm ESk-means algorithm

---

Inputs: A dataset $X = \{x_1, x_2, ..., x_N\}$, the Euclidian distance function $\| \, \|$.

Output: Optimal cluster scheme $C = \{C_1, C_2, ..., C_c\}$.

Step 1: Initialize the parameters related to k-means algorithm $c_{max} = c$, $c_{min} = 2$.

Step 2: Apply the k-means algorithm.

Step 3: If convergence, then go to step 3, else go to step 2.

Step 4: Calculate $SepCmp$ measure

Step 5: Repeat

   Apply splitting-based procedure for obtain the candidate cluster scheme.

   Increase the cluster number $c \leftarrow c + 1$.

   Calculate the $SepCmp$ measure for the new clusters.

  Until $c = c_{max}$

Step 6: Store $SepCmp$ measure which value is maximal.

---

## 5. Experimentation

### 5.1. *Data of Experimentation*

To validate the proposed approaches in determination of the number of clusters and clustering quality evaluation, three data sets are used among the various data files placed at the disposal of the artificial training community by the University of California with Irvine (UCI) [13] as well as an artificial data file coming from the benchmark "Concentric":

- Iris: this data set contains 3 clusters. Each cluster refers to a type of flower of the Iris: Setosa, Versicolor or Virginica. Each cluster contains 50 patterns of which each one has 4 components. The 1st cluster (Setosa) is linearly separable compared to the others, the two other clusters (Versicolor and Virginica) overlapped.

- Wine: this data set counts the results of a chemical analysis of various wines produced in the same area of Italy starting from various type of vines. The concentration of 13 components is indicated for each of the 178 analyzed wines which are distributed as follows: 59 in the 1st cluster, 71 in the 2nd cluster and 48 in the 3rd cluster.

- Diabetes: this data set counts the results of an analysis concerning the diabetes, carried out on 7 donors to diagnose the disease. The size of this data set is equal to 768 patterns distributed in two clusters. 500 for the 1st cluster and 268 for the 2nd cluster. Patterns are with 8 dimensions.

- Concentric: this data set is artificial and rather complex. It contains 2500 patterns, 1579 for the 1st cluster and 921 for the 2nd cluster. The 1st cluster is inside the second. Patterns are with 2 dimensions.

## 5.2. *Comparative Study*

In this sub-section, we propose to evaluate the proposed quality measures and clustering algorithms.

### 5.2.1. *Clustering Quality Evaluations*

In this sub-section, we propose to evaluate the proposed quality measures and clustering algorithms.

We present a comparative study of the proposed clustering algorithms via quality evaluation measures. We represent only some iterations for each data set

Table 1. Results of the evaluation of the EMk-meanss and ESk-means algorithms via Iris dataset.

| Iterations | Algorithm | $MSE$ | $I_{Dum}$ | $I_{DB}$ | $SepCmp$ |
|---|---|---|---|---|---|
| ... | **...** | ... | ... | ... | ... |
| | **...** | ... | ... | ... | ... |
| 6 | EMk-means | 0.41 | 0.27 | 0.57 | 0.34 |
| | ESk-means | 0.65 | 0.53 | 0.71 | 0.31 |
| 5 | EMk-means | 0.32 | 0.19 | 0.31 | 0.52 |
| | ESk-means | 0.48 | 0.39 | 0.48 | 0.43 |
| 4 | EMk-means | 0.27 | 0.14 | 0.20 | 0.65 |
| | ESk-means | 0.29 | 0.25 | 0.29 | 0.61 |
| 3 | EMk-means | 0.11 | 0.11 | **0.06** | **0.74** |
| | ESk-means | 0.21 | 0.15 | **0.11** | **0.69** |
| 2 | EMk-means | **0.09** | **0.02** | 0.14 | 0.61 |
| | ESk-means | **0.12** | **0.09** | 0.21 | 0.59 |

As shown in table 1, only $I_{DB}$ and $SepCmp$ measures determined the optimal number of clusters with respective values (0.11) and (0.69).

This is explained by the fact why the maximum value (resp. minimal) of $SepCmp$ measure is associated to the optimal number of clusters.

For the two proposed algorithms (EMk-means and ESk-means), $MSE$ and $I_{Dum}$ measures can not determine this number.

Table 2. Results of the evaluation of the EMk-meanss and ESk-means algorithms via Wine dataset.

| Iterations | Algorithms | $MSE$ | $I_{Dum}$ | $I_{DB}$ | $SepCmp$ |
|---|---|---|---|---|---|
| ... | **...** | ... | ... | ... | ... |
| | **...** | ... | ... | ... | ... |
| 7 | EMk-means | 0.96 | 0.98 | 0.91 | 0.34 |
| | ESk-means | 1.01 | 1.09 | 1.06 | 0.31 |
| 6 | EMk-means | 0.79 | 0.79 | 0.64 | 0.87 |
| | ESk-means | 0.84 | 0.86 | 0.89 | 0.79 |
| 5 | EMk-means | 0.41 | 0.51 | 0.49 | 1.02 |
| | ESk-means | 0.59 | 0.65 | 0.54 | 0.88 |
| 4 | EMk-means | 0.28 | 0.31 | 0.19 | 1.25 |
| | ESk-means | 0.32 | 0.43 | **0.22** | 1.03 |
| 3 | EMk-means | 0.13 | 0.17 | **0.15** | **1.31** |
| | ESk-means | 0.19 | 0.28 | 0.35 | **1.19** |
| 2 | EMk-means | **0.04** | **0.01** | 0.21 | 1.15 |
| | ESk-means | **0.10** | **0.03** | 0.51 | 0.99 |

As shown in table 2, only $SepCmp$ measure could determine the optimal number of clusters via the two proposed clustering algorithms. With $I_{DB}$ measure, the optimal number of clusters is obtained only with the EMk-means algorithm with the (0.15). $MSE$ and $I_{Dum}$ measures did not determine this number with the two algorithms.

Table 3. Results of the evaluation of the EMk-meanss and ESk-means algorithms via Diabète dataset.

| Iterations | Algorithms | $MSE$ | $I_{Dum}$ | $I_{DB}$ | $SepCmp$ |
|---|---|---|---|---|---|
| ... | **...** | ... | ... | ... | ... |
| | **...** | ... | ... | ... | ... |
| 9 | EMk-means | 1.13 | 1.06 | 0.99 | 0.48 |
| | ESk-means | 1.21 | 1.15 | 1.07 | 0.43 |
| 8 | EMk-means | 1.05 | 0.91 | 0.89 | 0.53 |
| | ESk-means | 0.98 | 0.98 | 0.94 | 0.40 |
| 7 | EMk-means | 0.64 | 0.83 | 0.77 | 0.66 |
| | ESk-means | 0.87 | 0.94 | 0.79 | 0.63 |
| 6 | EMk-means | 0.46 | 0.69 | 0.64 | 0.85 |
| | ESk-means | 0.75 | 0.73 | 0.63 | 0.74 |
| 5 | EMk-means | 0.43 | 0.46 | 0.39 | 1.16 |
| | ESk-means | 0.54 | 0.57 | 0.46 | 0.87 |
| 4 | EMk-means | 0.29 | 0.38 | 0.37 | 1.41 |
| | ESk-means | 0.39 | 0.39 | 0.29 | 1.08 |
| 3 | EMk-means | 0.23 | 0.19 | 0.21 | 1.59 |
| | ESk-means | **0,25** | 0.24 | **0.24** | 1.22 |
| 2 | EMk-means | **0.20** | **0.11** | **0.18** | **1.63** |
| | ESk-means | 0.27 | **0.17** | 0.31 | **1.45** |

As shown in table 3, with the EMk-means algorithm, all measures ($MSE$, $I_{Dum}$, $I_{DB}$ and $SepCmp$) determine the optimal number of clusters with the respective values (0.20), (0.11), (0.18), (1.63).

With the ESk-means algorithm, only measures $I_{Dum}$ et $SepCmp$ determine this number with the respective values (0.17) and (1.45).

Table 4. Results of the evaluation of the EMk-meanss and ESk-means algorithms via Concentric dataset.

| Iterations | Algorithms | $MSE$ | $I_{Dum}$ | $I_{DB}$ | $SepCmp$ |
|---|---|---|---|---|---|
| … | **…** | … | … | … | … |
| | **…** | … | … | … | … |
| 10 | EMk-means | 1.43 | 1.35 | 1.84 | 0.98 |
| | ESk-means | 1.57 | 1.11 | 1.49 | 0.98 |
| 9 | EMk-means | 1.25 | 1.06 | 1.56 | 1.15 |
| | ESk-means | 1.44 | 0.93 | 1.23 | 0.89 |
| 8 | EMk-means | 1.01 | 0.91 | 1.12 | 1.30 |
| | ESk-means | 1.20 | 0.87 | 0.92 | 1.08 |
| 7 | EMk-means | 0.65 | 0.84 | 0.98 | 1.52 |
| | ESk-means | 0.89 | 0.65 | 0.79 | 1.16 |
| 6 | EMk-means | 0.65 | 0.69 | 0.73 | 1.74 |
| | ESk-means | 0.81 | 0.54 | 0.61 | 1.65 |
| 5 | EMk-means | 0.48 | 0.51 | 0.69 | 2.13 |
| | ESk-means | 0.59 | 0.39 | 0.46 | 2.08 |
| 4 | EMk-means | 0.27 | 0.36 | 0.45 | 2.59 |
| | ESk-means | 0.31 | 0.28 | 0.33 | 2.45 |
| 3 | EMk-means | 0.14 | 0.21 | 0.32 | 2.92 |
| | ESk-means | **0.17** | **0.16** | **0.19** | **2.83** |
| 2 | EMk-means | **0.12** | **0.19** | **0.21** | **3.09** |
| | ESk-means | 0.20 | 0.28 | 0.24 | 2.82 |

As shown in table 4, with the EMk-means algorithm, all measures ($MSE$, $I_{Dum}$, $I_{DB}$ and $SepCmp$) determine the optimal number of clusters with the respective values (0.12), (0.19), (0.21), (3.09).

With the ESk-means algorithm, all used measures could not determine this number.

However, the values obtained by $SepCmp$ measure via the ESk-means algorithm for 2 and 3 number of clusters are respectively (2.83) and (2.82).

These two values are very close and almost equal.

Based on these results, we can say that the proposed clustering algorithms always give more powerful results for any type of dataset.

## 5.2.2. *Distribution Uniformity*

In this sub section, we propose to evaluate the two proposed clustering algorithms in terms of distribution of data in the various clusters.

Table 5. Results of the evaluation of the distributions of data via EMk-means and ESk-means.

| Dataset | Optimal number of clusters | Algorithms | Data distribution | | |
|---|---|---|---|---|---|
| | | | Realy | Obtained | Similarity |
| Iris | 3 | EMk-means | 50, 50, 50 | 50, 48, 52 | 2.6% |
| | | ESk-means | | 50, 47, 53 | 4.0% |
| Wine | 3 | EMk-means | 59, 71, 48 | 62, 71, 47 | 2.4% |
| | | ESk-means | | 62, 70, 46 | 3.4% |
| Diabète | 2 | EMk-means | 500, 268 | 449, 269 | 0.28% |
| | | ESk-means | | 446, 272 | 1.14% |
| Concentric | 2 | EMk-means | 1579, 921 | 1571, 929 | 0.69% |
| | | ESk-means | | 1567, 933 | 1.03% |

As shown in table 5, In the case of Iris data set, the similarity between really and obtained data distribution with algorithm EMk-means (resp. ESk-means) is equal to 2,6% (resp. 4,0%).

In the case of Wine data set, the similarity between really and obtained data distribution with algorithm EMk-means (resp. ESk-means) is equal to 2,4% (resp. 3,0%).

In the case of Diabetes data set, the similarity between really and obtained data distribution with algorithm EMk-means (resp. ESk-means) is equal to 0,28% (resp. 1,14%).

In the case of the Concentric data set, the similarity between really and obtained data distribution with algorithm EMk-means (resp. ESk-means) is equal to 0,69% (resp. 1,03%).

Compared to the size of the used data sets, we can conclude that these percentages are very satisfactory if the number of classes is unknown for the users.

We also note that the obtained results while following the merge-based strategy are more interesting than those are given by the splitting-based strategy.

## 5.3. *About Complexity*

The theoretical complexity of $SepCmp$ measure is based on the complexity of two terms defined by $Sep$ and $Cmp$ measures.

For a data set $X = \{x_i : i = 1,...,N\} \subset R^M$, $c$ is the initial number of cluster, theoretical complexity of $Sep$ measure is about $o(NMc)$ while the complexity of $Cmp$ measure is about $o(NMc^2)$. Then, The complexity of $SepCmp$ is about $o(NMc^2)$.

Usually, $M \prec\prec N$, therefore the complexity of this measure for a specific cluster scheme is about $o(Nc^2)$.

We present in table 6 a study of the theoretical complexity of the proposed clustering algorithms:

Table 6. Study of the complexity of the proposed clustering algorithms.

| Algorithms | k-means | EMk-means | ESk-means |
|---|---|---|---|
| Theoretical complexity | $o(Nc)$ | $o(Nc^2)$ | $o(Nc^2)$ |

As shown in table 6, the complexity of these algorithms is square proportional to the size of the data and to the maximum number of clusters.

## 6. Conclusion

A conclusion section is not required. Although a conclusion may review the main points of the paper, do not replicate the abstract as the conclusion. A conclusion might elaborate on the importance of the work or suggest applications and extensions.

The majority of clustering algorithms often run up against the problem of the optimal number of clusters to generate and therefore the clustering quality evaluation of obtained clusters. A solution consists in making iterations until obtaining a satisfying number of clusters. Each iteration tries to minimize a quality measure called validity index.

This quality is judged in general on the basis of two contradictory criteria. The first supposes that the generated clusters must be most various possible to each other with respect to certain characteristics, and the second requires that each cluster have to be most homogeneous possible with respect to these characteristics.

While inspiring by published algorithms, we have proposed mapping rules for generalizing these last.

Our major contributions are as follows:

- Definition of global quality measures of a partition generated by a clustering algorithm while basing on compactness and separation measures,
- Definition of local quality measures to identify the "worst" clusters to be merge or split.
- Binarization of the merging and splitting processes.

Towards Fuzzy-Hard Clustering Mapping Processes

- Modeling and implementation of two clustering approaches implementing the new proposed measures and processes. The first proposed approach is based on the principle of merging. It starts with a maximum number of clusters and decreases it during various iterations. The second proposed approach is based on the principle of splitting. It starts with a minimum number of clusters which is incremented during its execution. For the two approaches, the basic idea consists in determining the optimal number of clusters, by global and local evaluations, of the obtained partition.

Proposed measures, processes and approaches are exploited successfully for various data sets.

Future work relate to primarily the test of these approaches on large data sets. Indeed, we must simplify this operation by a data sampling while allowing a better evaluation of this last.